# Quantum Vacuum and a Matter - Antimatter Cosmology

## Frederick Rothwarf [1,2] and Sisir Roy [3,4]

1 *Department of Physics, George Mason University, Fairfax, VA 22030 USA*

2 *Magnetics Consultants,* 11722 *Indian Ridge Road, Reston, VA 20191.,USA*

3 *Center for Earth Observing and Space Research, College of Science, George Mason Univ., Fairfax, VA 22030 USA*

4 *Physics and Applied Mathematics Unit, Indian Statistical Institute, Calcutta, INDIA*

*2* e-mail: frothw@ieee.org

*3* e-mail: sisir@isical.ac.in

**Abstract**
A model of the universe as proposed by Allen Rothwarf based upon a degenerate Fermion fluid composed of polarizable particle-antiparticle pairs leads to a big bang model of the universe where the velocity of light varies inversely with the square root of cosmological time, t. This model is here extended to predict a decelerating expansion of the universe and to derive the Tully-Fisher law describing the flat rotation curves of spiral galaxies. The estimated critical acceleration parameter, $a_{Ro}$, is compared to the experimental, critical modified Newtonian Dynamics (MOND) cosmological acceleration constant, $a_o$, obtained by fitting a large number of rotation curves. The present estimated value is much closer to the experimental value than that obtained with other models. This model for $a_R(t)$ allows the derivation of the time dependent radius of the universe R(t) as a function of red shift z, R(z). Other cosmological parameters such as the velocity of light, Hubble's constant, the Tully-Fisher relation, and the index of refraction of the aether can also be expressed in terms of z. R(z) is compared with the statistical fitting for Veron-Cetty data (2006) for quasar red shifts and good agreement is found. This model also determines the time and/or z dependence of certain electromagnetic parameters, i.e., the permittivity $\varepsilon_v(t)$; the permeability $\mu_v(t)$; and index of refraction *n*(t) of free space. These are found to be useful in various cosmological theories dealing with light passing through media in motion.

**I - Introduction**
For a long time the concept of an aether as a medium for the propagation of electromagnetic waves has been discredited, even though Maxwell's equations were originally derived based upon the assumption of an aether. Today the need for something like aether is acknowledged in physics by invoking terms such as "quantum vacuum," "vacuum fluctuations," or "zero-point fluctuations." In fact Grossing has recently shown how the Schrodinger equation can be derived by invoking such zero-point fluctuations.[1]

Allen Rothwarf[2] has reviewed the objections to an aether and concluded that it was indeed needed to explain many problems in physics such as wave-particle duality; the nature of spin; the derivation of Hubble's law; electric fields; Zitterbewegung; inflation in cosmology; the arrow of time; the Pauli exclusion principle; the nature of the photon; neutrinos; redshifts; and several other ideas. In that paper he referred to previous aether models but finally chose to explore a model based upon a degenerate Fermion fluid composed of polarizable particle-antiparticle pairs, e.g., electron-positron pairs. This leads to a big bang model of the universe, where the velocity of light varies inversely with the square root of cosmological time, t.
He was motivated in part by Dirac's 1951 letter to Nature titled "Is There an Aether?"[3] in which Dirac showed that the objections to an aether posed by relativity were removed by quantum mechanics, and that in his reformulation of electrodynamics the vector potential was a velocity.[4] Dirac concludes his letter with "We have now the velocity at all points of space-time, playing a fundamental role in electrodynamics. It is natural to regard it as the velocity of some real physical thing. Thus with the new theory of electrodynamics we are rather forced to have an aether."

In this paper we expand on the electron-positron aether model to obtain its further cosmological implications. The model is used to determine the time dependence of certain fundamental constants, i.e., the velocity of light, c(t); the permittivity $\varepsilon_v(t)$; the permeability $\mu_v(t)$; and index of refraction of free space $n(t)$. One can also find both the radius of the universe R(t) and the redshift z(t) as a function of time. z(t) is shown to be an exponential function of R, which in turn is related to the apparent luminosity, **m**, of stars. A statistical analysis to determine the best fit to log z versus m of data for nearly 49,000 quasars (Vernon-Cetti Catalogue (2003)[5] also gave an exponential curve for z versus m, which closely matches that derived from the aether model.

Milgrom[6] proposed an ad hoc modification of Newton's law of gravity or inertia, known as modified Newtonian dynamics (MOND) to eliminate the need for dark matter. MOND explains remarkably well the systematic properties of spiral or elliptical galaxies and predicts in detail the observed rotation curves of such galaxies. It does so by invoking only one critical parameter, $a_o$, the MOND cosmological acceleration constant. <u>In the MOND model $a_o$ is considered to be a universal deceleration constant much as c is a universal velocity constant.</u> Fitting the rotation curves for a large number of galaxies[7,8] gives a value $a_o = 1.2 \times 10^{-10}$ m/sec$^2$, which Milgrom estimated to be $\sim c_o H_o/6$,[8] where $c_o$ and $H_o$ are the present values of the velocity of light and Hubble's constant, respectively.

It is shown in the present work that the aether model gives a theoretical basis for the MOND model and yields an acceleration parameter which closely matches the experimental value for $a_o$. First, we give a brief review of the Rothwarf aether model in section II. Then in section III we extend this model so as to calculate the acceleration parameter; the distance-redshift relation; and the time dependence of permittivity, permeability and index of refraction. A brief review of Modified Newtonian Dynamics (MOND) and a critical analysis of it are given in section IV. In section V the Tully-Fisher law is derived by using the aether model. The implications of our results are discussed in section VI.

**II – Review of the Rothwarf Aether Model**
Rothwarf[2] proposed a model of the vacuum based upon the concept of a degenerate Fermi fluid, which consists of particles-antiparticles, e.g., electrons-positrons in a negative energy state relative to the null state or true vacuum. In this model the aether condenses soon after the big bang as a plasma of particles that obey Fermi-Dirac statistics. The particles will have a velocity, and hence the region containing these particles will expand even after new particle-antiparticle production ceases. *Therefore, in this picture the expansion of the aether replaces space-time and the aether model should give results equivalent to those derived by relativity models.*

The universal feature of a degenerate Fermi fluid model is the existence, due to the Pauli Exclusion Principle, of a highest energy (at zero temperature) for the particles called the Fermi energy level. The velocity associated with the particles having this highest energy is called the Fermi velocity. The Fermi velocity can be expressed as

$$v_F = hk_F/2\pi m = (3\pi^2 n_e)^{1/3} h/2\pi m, \qquad (1)$$

where h is Planck's constant, m is the mass of the electron (positron), $k_F$ is the wavevector of electrons at the Fermi energy and $n_e$ is the density of the electrons (positrons). In Rothwarf's model, the Fermi velocity was equated with the speed of light c. The reason behind this assumption is that the excitations in such a system travel at velocities limited by $v_F$. Assuming, *h* and m as time independent, he came to the following interesting conclusions:

c depends on the density of the particles (antiparticles) of the aether and is not a quantity that we must accept as given. Furthermore, as $n_e$ *decreases* with time due to the expansion of the aether, c can be considered as decreasing function of cosmological time.

The assumption that $v_F = c$ has direct implications[2]. As the aether expands and the big bang cools, particle production ceases at some point and leaves a total number of electrons (positrons) $N_o$ in the aether. If $R(t)$ is the radius of the aether (i.e., the radius of the universe) then $n_e = N_o/(4/3 \pi R^3(t))$. Within the aether where $c(t)$ is the highest speed of the particles, some will always be crossing the outer boundary with the true vacuum, and thus expanding $R(t)$. The rate of expansion of the aether R' will be proportional to $c(t)$. Rothwarf assumed that one can take $R'(t) = \alpha c(t)$, where $\alpha \sim \frac{1}{2}$, and rewrite Eqn (I) as

$$c(t) = R'(t)/\alpha = h/2m \{ 3\pi^2 N_o/(4\pi/3) \}^{1/3}, \qquad 1/R(t) = c_o R_o / R(t) \qquad (2)$$

where $c_o$ is the present speed of light, $R_o$ is the present radius of the universe, and $R'(t)$ is the time derivative of $R(t)$. This gives the differential equation

$$R(t)R'(t) = \alpha c_o R_o \qquad (3)$$

where $\alpha$ is a geometrical factor $\sim 1/2$. Using $R(t)R'(t) = (\frac{1}{2}) dR^2(t)/dt$, the solution of this equation can be written as

$$R(t) = [2\alpha c_o R_o(t - t_i) + R_i^2]^{1/2} \qquad (4a)$$

For $t \gg t_i$, this becomes

$$R(t) = [2\alpha c_o R_o t]^{1/2} = R_o (t/t_o)^{1/2}, \quad \text{where} \quad R_o = 2\alpha c_o t_o \qquad (4b)$$

and from Eqn (2)

$$c(t) = (c_o R_o/2\alpha)^{1/2} \ t^{-1/2} = c_o (t/t_o)^{-1/2} \qquad (4c)$$

where $t_i$ is the time at which particle production ceased and $R_i$ is the radius of the universe at that time. Since the present time $t_o \gg t_i$ and $R_o \gg R_i$, Eqn (4) predicts that the universe is expanding with a $t^{1/2}$ time dependence. Also, from Eqn (4), and the second form of Eqn (2), we see that the speed of light is deceasing as $t^{-1/2}$. The time dependence of $R(t)$, given by Eqn (4) is nearly identical to that found from relativity for an Einstein-De Sitter universe dominated by radiation. Moreover, another result that arises naturally from the aether fluid is Hubble's law. This law, which was deduced from experimental observations, states that the farther away from us a galaxy is, the greater is its velocity away from us. Mathematically that is,

$$\rho' = -H\rho \qquad (5)$$

$H$ is the Hubble's constant and $\rho = [R_o - R(t)]$ is the distance from earth to the observed galaxy. Making use of the continuity equation for fluid flow, Rothwarf was able to derive a time dependent expression for Hubble's law with $H(t)$ given by:

$$H(t) = R'(t)/R(t) = -\rho'/\rho = \alpha c_o R_o / R^2(t) = 1/2t, \text{ or } H(t) = H_o (t_o/t) \qquad (6)$$

where $H_o$, $c_o$, $R_o$, and $t_o$ are the present values of Hubble's constant, the speed of light, the radius of the universe (aether) and cosmological time respectively. For the present time $H_o = 1/2t_o$, which is one half the presently accepted value.

**III – An Extension of the Aether Model**
The cosmological aspects of the model can be extended to address the acceleration of the aether. This can then be used to derive a relation between redshift as a function of time, z(t) and in turn relate that to R(t) to obtain R as a function of z., Furthermore, the model is used to determine the time dependence of permittivity, permeability, and index of refraction for the aether.

## A – Acceleration of the Aether

To find the aether acceleration one differentiates $R'(t)$ in Eqn (2) and gets

$$R''(t) = -[\alpha c_o R_o / R^2(t)] \, R'(t) = - R'(t) / 2t \quad \text{for } t \gg t_i \text{ and } R \gg R_i \tag{7}$$

This can be rewritten using Eqns (2) and (6) and letting $a_R(t) = R''(t)$ as

$$a_R(t) = R''(t) = -\alpha c(t)/2t = -\alpha c(t) H(t) = -(\alpha c_o / 2 t_o) x^{-3/2} = -a_{Ro} x^{-3/2} \tag{8}$$

The term $a_R(t)$ is the aether acceleration parameter, which in fact shows that the aether is decelerating. For the present era where $c_o = 3.00 \times 10^8$ m/s; $t_o = 13.7 \times 10^9$ years = $4.33 \times 10^{11}$ s; and assuming $\alpha = 1/2$ one calculates that $a_{Ro} = -1.7 \times 10^{-10}$ m/s$^2$. However, we have reason to question the presently accepted value of $t_o = 13.7 \times 10^9$ years. This question will be reserved for later discussion. The significance of $a_{Ro}$ will also be discussed below when MOND issues are considered.

All of the above quantities $R(t)$, $c(t)$, $H(t)$, and $a_R(t)$ can be divided by their present values to give a set of reduced variables as a function of $x = t/t_o$ as follows:

$$\mathbf{r} = R(t)/R_o = x^{1/2}; \quad \mathbf{c} = c(t)/c_o = x^{-1/2}; \quad \mathbf{h} = H(t)/H_o = x^{-1}; \quad \mathbf{g} = a_R(t)/a_{Ro} = x^{-3/2} \tag{9}$$

These curves are shown in Figure 1.

## B – Redshift Considerations

As one looks back in time from the present time $t_o$, when the radius of the universe is $R_o$, toward some distant galaxy at time t and cosmological radius R, the light detected is red-shifted by a factor of $z = \Delta\lambda/\lambda$ with respect to that on earth. The distance from earth to the galaxy is $\rho = [R_o - R(t)]$. One can write the differential expression relating the change in velocity $\rho''$ with which the galaxy is receding to the change in z with time, z':

$$\rho'' = c(t) \, z'. \tag{10}$$

But $\rho'' = -a_R(t) = -[-\alpha c(t)/2t]$, so that Eqn (10) can be integrated to give

$$z(t) = (\alpha/2) \ln(t_o/t) = (\alpha/2) \ln(1/x) \quad \text{or} \quad x = e^{-(2/\alpha) z} \tag{11}$$

This function is plotted in Figure 2 for $\alpha = \frac{1}{2}$ and $\alpha = 1$. All of the reduced variables in Eqn (9) can now be expressed as a function of the red shift z; e.g.,

$$\mathbf{r} = x^{1/2} = e^{-z/\alpha}; \quad \mathbf{c} = x^{-1/2} = e^{z/\alpha}; \quad \mathbf{h} = x^{-1} = e^{(2/\alpha)z}; \quad \mathbf{g} = x^{-3/2} = e^{(3/\alpha)z} \tag{12}$$

These functions are shown in Figure 3 for $\alpha = \frac{1}{2}$. From Eqn (12) where $R(t)/R_o = x^{1/2}$, we have $R(z) = R_o \, e^{-z/\alpha}$. This in turn yields the distance from earth $\rho(z)$:

$$\rho(z) = R_o \, [1 - e^{-(z/\alpha)}], \quad \text{where} \quad R_o = 2\alpha c_o t_o \tag{13}$$

Since data for galaxies and quasars are often plotted as log z versus apparent magnitude, m, and m can be related to $\rho(z)$, one can compare Eqn (13) to the statistical fit to actual data taken by various surveys such as the Veron-Cetty Catalogue for Quasars.[5,35] The comparison will depend upon the choice of $\alpha$.

## C – Time Dependence of Permittivity, Permeability, and Index of Refraction

Recently Puthoff[9] published a Polarizable-Vacuum approach to General Relativity (GR) in which the basic postulate is that the polarizability of the vacuum in the vicinity of mass differs from its asymptotic far-field value. Thus, he proposed that for the vacuum itself

$$D = \varepsilon E = K\varepsilon_o E \tag{14}$$

$K$ is the altered dielectric constant of the vacuum (assumed to be a function of position in his formulation), due to changes in vacuum polarizability (GR induced).

In the present paper $K$ is considered as a function of cosmological time, $K(t)$. We consider that the expected polarizability of the vacuum (aether) will be changing as the density of the aether decreases with the expansion of the universe. This is consistent with the assumption of this model which states that the number of electron-positron pairs in the universe remains constant after the end of the inflationary phase of the big bang.

We begin our analysis by considering the fine structure constant, $\alpha$, that governs electromagnetic interactions, i.e.,

$$\alpha = e^2 / (2\varepsilon_o hc_o) \tag{15}$$

$c_o = (\mu_o \varepsilon_o)^{-1/2}$ in the aether model considered here. In the present case, e, and h are taken as constants; $c_o$, $\varepsilon_o$, and, $\mu_o$ are the present values of the speed of light, the vacuum permittivity, and the vacuum permeability, respectively. Taking into consideration that $\varepsilon_o$ is expected (with a time-varying polarizability) to change to $\varepsilon(t) = K(t) \varepsilon_o$, the fine structure constant can be rewritten as,

$$\alpha = e^2 / (2\, K(t)\, \varepsilon_o hc(t)) \tag{16}$$

There is reason to believe that $\alpha$ has not varied significantly with time since the end of inflation. The observations of quasar absorption spectra by Webb et al[10] showed that $\alpha$ was slightly lower in the past, with $\Delta\alpha/\alpha_o = -0.72 +/- 0.18 \times 10^{-5}$ for $0.5 < z < 3.5$. Analyzing geological constraints, imposed on a natural nuclear fission event at Oklo, Darmour and Dyson[11] concluded that $\Delta\alpha/\alpha_o$ over the past 1.5 billion years has been $< 5.0 \times 10^{-17}$ yr$^{-1}$. Therefore, with some confidence one can substitute Eqn (4c) into Eqn (16) to obtain from $\Delta\alpha/\alpha_o = [1 - \alpha(t)/\alpha_o] \sim 0$ that

$$K(t) = (t/t_o)^{1/2}. \tag{17}$$

Then the permittivity is given by

$$\varepsilon(t) = K(t)\, \varepsilon_o = \varepsilon_o (t/t_o)^{1/2}. \tag{18}$$

Now rewriting the expression for c(t) as $c(t) = c_o (t/t_o)^{-1/2} = [\mu(t)\varepsilon(t)]^{-1/2}$ where $c_o = (\mu_o \varepsilon_o)^{-1/2}$, one solves for $\mu(t)$ using Eqns (17) and (18) to obtain

$$\mu(t) = \mu_o (t/t_o)^{1/2}. \tag{19}$$

The index of refraction, $n$, for the aether is given by

$$n(t) = c_o / c(t) = (t/t_o)^{1/2} = x^{1/2}. \tag{20}$$

The Rothwarf aether model shows that $\mu(t)$, $\varepsilon(t)$ and $n(t)$ all increase with the square root of cosmological time. The index of refraction, n(t) can be expressed as a function of z by using Eqn (11) to give $n(t) = e^{-z/\alpha}$, which is the same function as that for the radial expansion of the universe. The expression $(c(t)/c_o)^2 = 1/n^2(t) = (t/t_o)^{-1} = e^{(2/\alpha)z}$ varies inversely with time and is used below in discussion in connection with the work of Leonhardt and Piwnicki[12,13] and Spavieri[14] concerning the optical behavior of flowing dielectrics.

## IV – Modified Newtonian Dynamics (MOND)

Milgrom[6a,b,c] proposed an ad hoc modification of Newton's law of gravity, known as modified Newtonian dynamics (MOND), to eliminate the need for dark matter. MOND modifies Newtonian dynamics in the region of very low acceleration. Newton's law has the gravitational force proportional to $r^{-2}$, where r is the radial position, whereas MOND uses an $r^{-1}$ law which fits the data very well in extragalactic regions where dynamical accelerations are small, see Begeman, et al.[7] It does so by invoking only one critical p*arameter, $a_o$,* the MOND cosmological acceleration constant. Again we note <u>$a_o$ is considered to be a universal deceleration constant much as c is a universal velocity constant.</u> The fitting of rotation curves of a large number galaxies gives a value $a_o = 1.2 \times 10^{-10}$ m/sec$^2$, which Milgrom estimated was ~ $c_oH_o$, to within a factor of about six, where $c_o$ is the present velocity of light and $H_o$ is the present value of Hubble's constant.

Much theoretical work has been done trying to derive $a_o$ from cosmological models[19-30] (see these works for extensive reviews of such efforts). Both Beckenstein[22] and Carmeli[23-26] have formulated modifications and extensions to Einstein's general theory of relativity to derive the constant $a_o$. Carmeli's model takes into account the Hubble expansion, which imposes an additional constraint on the motion of particles. He postulates that the usual assumptions in obtaining Newton's gravitational law from general relativity are insufficient, so that gases and stars in the arms of spiral galaxies must also be governed by Hubble flow. As a result a universal constant $a_{oC}$ is introduced as the minimum acceleration in the cosmos. This differs from the Milgrom value.

With his theory Carmeli[23] provided a successful derivation of the Tully-Fisher law. Hartnet[27] using the Carmeli metric found the relationship between the fourth power of the galaxy's rotation speed ($v_c$) and its mass (M) to be

$$v_c^4 = (2/3) \, a_o GM \tag{21}$$

G is the gravitational constant. Eqn (21) can be re-written as $v_c^4 = (2/3) \, a_o \, (GM/r^2) \, r^2$ and thus the positive square root is $v_c^2 = [(2/3) \, a_o \, g_N]^{1/2} r$, where $g_N$ is the Newtonian gravitational acceleration. This latter expression is consistent with Milgrom's phenomenological approach in the low acceleration limit, which gave $v_c^2 = [a_o g_N]^{1/2} r$. It should be noted that this latter expression was used to fit experimental data to obtain $a_o$.[8] Thus, the Carmeli theory gives the MOND constant to be $a_{oC} = (2/3) a_o$. Hartnett[27,28] has shown that the Carmeli metric correctly describes spiral galaxy rotation curves and gravitational lensing without the need for dark matter.

As was noted above in the Introduction, Milgrom had originally estimated that $a_{oM} \sim c_o H_o$, to within a factor of about six, whereas in the Carmeli theory $a_{oC} = c_o H_o$, and the acceleration term in the aether model was shown to be $a_{oR} = \alpha c_o H_o$. Below we show that it is more than just a coincidence that the aether model gives an acceleration parameter very similar to the MOND parameter result obtained by Carmeli using a modification of general relativity theory, namely, $a_{oR} = \alpha \, a_{oC}$. We do this by obtaining the Tully-Fisher law with use of the aether model.

## V – Derivation of the Tully-Fisher Law Using the Aether Model

As Carmeli notes,[23] the motion of a star around its galaxy must also experience the expansion of the universe in addition to the Newtonian gravitational attraction to the galaxy's center of mass. The expansion of the universe changes the distance between the star and the center of the galaxy and consequently changes the circular velocity of the star. Following Carmeli and Goldstein[34] we write an "effective potential" for the motion of a star in a central field as

$$V_{eff}(r) = - GM/r + L^2/2r^2 - a_R(t) \, r \tag{22}$$

L is the angular momentum per unit mass, and $a_R(t)$ is given by Eqn (8). The minimum value of Eqn (22) gives the condition for a stable circular orbit, i.e.,

$$dV_{eff}(r)/dr = 0 = GM/r^2 - L^2/r^3 - a_R(t) \tag{23}$$

$L = v_c r$, and $v_c$ is the rotational velocity. This leads to

$$v_c^2 = GM/r - a_R(t)\, r \qquad (24)$$

and

$$v_c^4 = (GM/r)^2 - 2GM\, a_R(t) + [a_R(t)\, r]^2 \qquad (25)$$

The first term on the right hand side of Eqn (25) is the Newtonian term which is negligible in the flat region of the galaxy rotation curve. The second one is the Tully-Fisher term. The third term is very small due to the smallness of $[a_R(t)]^2$. As was shown above $v_c^2$ can be expressed as

$$v_c^2 = [-2a_R(t)GM]^{1/2} = [-2a_R(t)g_N]^{1/2} r \qquad (26)$$

where $g_N$ is the Newtonian gravitational acceleration. Since $a_R(t) = -\alpha c(t)/2t = -\alpha c(t)H(t)$, the bracket is in fact a positive number. From Eqns (9), (11) and (12) we can now write Eqn (26) as

$$v_c^2 = [(\alpha c_o/t_o)\, e^{(3/\alpha)z}\, GM]^{1/2} = [2a_{Ro} e^{(3/\alpha)z}\, GM]^{1/2} \quad \text{or} \quad v_c = [2a_{Ro} e^{(3/\alpha)z}\, GM]^{1/4} \qquad (27)$$

Thus, we have shown that in fact the MOND term in the Tully-Fisher relation should depend upon the red shift z associated with a given galaxy. Many of the galaxies considered by Sanders and McGaugh[8] were at about the same distance from earth, $\rho$ = 15.5 Mpc. Using Eqn (11) one calculates z = 2.74 x 10$^{-3}$ for an $R_o$ = 2.83 x 10$^3$ Mpc. Thus for z ~ 0, $2a_{Ro}$ would correspond to the $a_o$ in the Milgrom model and (2/3) $a_{oC}$ for the Carmeli model, when one use Tully-Fisher analysis to fit an experimental "$a_o$" to the flat rotation curves of various galaxies.

**VI - Discussion –**
**A – Cosmological Time, Hubble's Constant, & the Aether Acceleration Term**
Recently the accepted value for the age of the universe, $t_o$, has been based upon the Wilkinson Microwave Anisotropy Probe (WMAP) measurements (reported by Spergel, et al[15]) which found $t_o$ = 13.7 x 10$^9$ years. However, more recent work has given cause to question that value. Bonanos, et al[16] have worked on the important galaxies M31 and M33 to calibrate the absolute extragalactic distance scale. They have found that previous distance measurements were too low by about 15 % for the first rung on the cosmological distance scale, so that the value previously accepted for Hubble's constant $H_o$ should be smaller by that amount. Furthermore, an extensive 15 year study using the *Hubble Space Telescope* by Sandage, et al[17] has found that $H_o$ = 62.3 kms$^{-1}$Mpc$^{-1}$, which in turn was 14 % smaller than the previously accepted value of 72 kms$^{-1}$Mpc$^{-1}$ given by Freedman, et al[18]. This Sandage value does not take account of the Bonanos result. For purposes of our analysis we will correct the Sandage result with the Bonanos correction to obtain **$H_o$ = 52.9 kms$^{-1}$Mpc$^{-1}$**.

In view of these concerns, we use the corrected value for $H_o$ in Eqn (6), where $H_o = 1/2t_o$, to determine $t_o$. We will use that value in our subsequent discussions. Thus, we find **$t_o$ = 9.24 x 10$^9$ yr or 2.92 x 10$^{17}$ s**. Conventional analysis uses $H_o = 1/t_{oc}$, which would then yield $t_{oc}$ = 18.5 x 10$^9$ yr or 5.84 x 10$^{17}$ s for the present age of the universe. The aether acceleration term given by Eqn (8) as $a_R(t) = -\alpha c(t)/2t$, now is recalculated for the present time as $a_{Ro} = -\alpha c_o/2t_o = -2.5 \times 10^{-10}$ m/s$^2$ for $\alpha = \frac{1}{2}$, while using the conventional value for $t_{oc}$ gives $a_{Ro} = -1.3 \times 10^{-10}$ m/s$^2$.

**B – MOND Considerations**
In the Introduction we claimed that with the aether model *the expansion of the aether replaces space-time and one should obtain results equivalent to those derived by relativity models.* In support of this contention we showed that the time dependence for the expansion of the universe *R(t)*, given by Eqn (4), is nearly identical to that found from relativity for an Einstein-De Sitter universe dominated by radiation. Moreover, another result that arises naturally from the aether fluid is Hubble's law. Furthermore, the theoretical deceleration of the aether expansion, $a_{Ro}$, corresponds very closely to the centripetal acceleration $a_{Co}$ given by Carmeli's relativity-based MOND model. Also we have been able to derive from the aether model the Tully-Fisher relationship and to show that it can be related to the redshift of a given galaxy. Thus, we argue that the aether deceleration supplies the local centripetal acceleration needed to account for the flat rotation curves observed for spiral galaxies.[8] It gives rather good agreement ($a_{Ro} = 2.5 \times 10^{-10}$ m/s$^2$ for $\alpha = \frac{1}{2}$) within a factor of two with the experimental value $a_o = 1.2 \times 10^{-10}$ m/sec$^2$ found by Begeman et al[7] and Sanders and McGaugh.[8] They note that the value of $a_o$ depends upon the value of $H_o$ (=75 kms$^{-1}$Mpc$^{-1}$) chosen by them to determine the distance scale for fitting their galaxy data. This "experimental" value is $\sim c_o H_o/6 \sim$ one–sixth the "theoretical" estimate, when one uses their assumed value of $H_o$, or $\sim c_o H_o/4$, when one uses $H_o = 52.9$ kms$^{-1}$Mpc$^{-1}$. Since we have chosen to calculate $a_{oR}$ by using $H_o = 52.9$ kms$^{-1}$Mpc$^{-1}$ in Eqn (8), we obtain a value of $a_{oR}$ closer to the experimental value. Nevertheless, we are off by only a factor of two not four. This good numerical prediction of the MOND constant gives important support for our claim that the modification of Newtonian dynamics as suggested by Milgrom[6a,b,c] can be due to the effect of the aether.

**C – Dark Matter**
Sanders and McGaugh[8] in an extensive analysis of thirty-eight spiral galaxies show that the MOND model gives a better fit to the flat rotation curves of spiral galaxies with fewer adjustable parameters than do various attempts to fit these curves with a dark matter model. In fact they use the same value of $a_o=1.2 \times 10^{-10}$ m/sec$^2$ found by Begeman et al[7] to analyze all the rotation curves and only adjust the mass to luminosity parameter for each galaxy. They conclude that the phenomenological MOND model gives a better description of the rotation curves of galaxies than does the dark matter model, but they worry about the lack of a good theoretical basis for MOND. In the present work we show the aether model gives good agreement with relativity based models for MOND. We wish to point out that in a sense the aether is equivalent to "dark matter" that pervades the universe. If one calculates the present density of electron-positron pairs from Eqn (1), as we showed above, one obtains a value $n_{eo} = 5.88 \times 10^{35}$ pairs/m$^3$. Rothwarf[2] pointed out that the real aether would be a mixture of various kinds of particle-antiparticle pairs, i.e., proton-antiproton and neutron-antineutron pairs in addition to the electron-positron pairs that have been considered here. If we assume these all have the same Fermi velocity, whose value is the speed of light, one can also calculate their respective densities from Eqn (1), since different fermion - antifermion pairs have different masses. Thus, one can speculate that each fermion type in the aether contributes its own "dark matter" component. For example, using Eqn (1) one can calculate that the present ratio of the proton-antiproton pair density, $n_{po}$, to the electron-positron pair density, $n_{eo}$ is given by

$$n_{po}/ n_{eo} = (m_p/m_e)^3 = 6.19 \times 10^9, \quad (28)$$

so that $n_{po} = 3.64 \times 10^{45}$ pairs/m$^3$. These are enormous numbers! Their implications will be discussed in another paper. In passing, we note the reduced time and z dependence of a given pair density, *q*, can be obtained from Eqns (1) (9) and (11) to yield

$$n_{\boldsymbol{q}}(t)/ n_{\boldsymbol{q}o} = [c(t)/c_o]^3 = x^{-3/2.} = e^{(3/\alpha)z} \quad (29)$$

This is the same function as the reduced acceleration term **g** presented above.

**D – Present Expansion of the Universe - Accelerating or Decelerating?**
The aether model presented here shows that the present expansion of the Universe is decelerating. This result conflicts with the current general belief that the present expansion of the Universe is accelerating driven by some hypothetical negative pressure called "dark energy." This belief is based upon high-redshift supernovae (SNe)Ia measurements[15] which cannot be explained by the decelerating Einstein- de Sitter model once in vogue before these results became available a few years ago. However, Vishwakarma[33] recently pointed out that with the present poor quality of the SNeIa data, the allowed parameter space is wide enough to allow decelerating models as well. He considered a particular example of the dark energy equation and was able to obtain a decelerating model consistent with recent high-redshift SNeIa data. He also noted that "if one takes into account the absorption of light by the intergalactic metallic dust that extinguishes radiation traveling over long distances, then the observed faintness of the extra-galactic SNeIa can be explained successfully in the framework of the Einstein-de Sitter model."

Vishwakarma further notes that while the best-fitting standard model to the SNeIa data indicates an accelerating expansion, there exist other low-density open models, which show a decelerating expansion that also fit the SNeIa observations reasonably well. Therefore, we believe that our decelerating aether expansion result is valid and also obviates the need for "dark energy."

**E – Redshift Considerations**
**(1) – Cosmic Time vs the Redshift**
In their paper ""The Cosmic Time in Terms of the Redshift," Carmeli, Hartnet, and Oliveira[30] derive the relation

$$t = 2t_o / [1 + (1 + z)^2] = 28 \text{ Gyr} / [1 + (1 + z)^2], \tag{30}$$

They assume $t_o \sim 14$ Gyr and compare this with a semi-empirical relationship given by Schwarzchild[31] as

$$t = 14 \text{ Gyr} / (1 + z)^{3/2}. \tag{31}$$

In Figure 4 we compare these two functions with our Eqn (11) plotted for $\alpha = \frac{1}{2}$. The aether model indicates much earlier cosmic times for a given value of z than do the other models.

**(2) - Statistical Analysis of Redshift vs Magnitude Data**
In a recent study entitled "Non-parametric Tests for Quasar Data and the Hubble Diagram," Roy, et al[35] have done a statistical analysis of the scatter plot of the truncated data for log redshift (z) vs apparent magnitude (**m**) in the case of quasars as compiled in the Veron-Cetty Catalogues (both in 2003 and 2006). Figure 5 shows the scatter plot of the Veron-Cetty (2006)[5] data for log z vs **m**. The idea of truncation is used here in the sense that the data ($z_i$, $m_i$) are observable, if they lie above the line log z = $am + b$, where $a = 3/7$ and $b = -64/7$. Of the 48,683 Veron-Cetty data points only 18 were below the line. A statistical fit to the data set is shown as the solid curve, which has a nearly exponential character.

The apparent magnitude **m**, which is a measure of luminosity of an object as it appears to us, and the redshift z of this object can be used to deduce the absolute or intrinsic magnitude M of the object, if one assumes the validity of a given cosmology, i.e., a distance-redshift relationship, $\rho(z)$. The relationship used is

$$\mathbf{m} - M = 5 \log [\rho(z)/10], \tag{32}$$

From our Equation (13)
$$\rho(z) = R_o [1 - e^{-(z/\alpha)}], \quad \text{where } R_o = 2\alpha c_o t_o.$$
Thus, we find
$$\mathbf{m} - M = 5 \log R_o /10 + 5 \log [1 - e^{-(z/\alpha)}]. \tag{33}$$

On substituting $t_o = 9.24$ Gyr or $2.92 \times 10^{17}$ s into the expression for $R_o$ one finds for $\alpha = \frac{1}{2}$ that $R_o = 8.77 \times 10^{22}$ km = $2.83 \times 10^3$ Mpc. Using this value for $R_o$ and the value of -22.5 for M from the statistical fit of Roy, et al[32] one obtains the dashed curve in Figure 5 for $\alpha = \frac{1}{2}$. This corresponds almost exactly with the Roy statistical fit for the range $-0.5 < \log z < 0.5$, i.e., for $0.32 < z < 3.2$. Such good agreement gives some confidence in the validity of the aether model and the choices of $t_o$ and $\alpha$ that were made.

**F – Electromagnetic Parameters**
It is clear from the analysis of the time dependence of various electromagnetic parameters like refractive index, permittivity and permeability, the value of c/$n$ varies with time i.e.

$$c(t)/n(t) = c_o/n^2 = c_o (t/t_o)^{-1} = c_o x^{-1} \tag{34}$$

It is worth mentioning that recently several authors (Leonhardt and Piwnicki[12,13] and Spavieri[14] - see references therein) considered the propagation of light in a moving non-dispersive dielectric medium. In the present framework, the aether like particle-antiparticle medium is time dependent.

Now if the light comes from a quasi stellar object (QUASAR) and travels for millions of years in a medium with time dependent electromagnetic parameters, we can rethink of the following results. Fresnel[35] in 1818 showed theoretically, by assuming an aether, that the speed of light in a uniform yet moving medium, $c_m$, of refractive index $n$ depends on the medium velocity $u$ and can be written as

$$c_m = c_0/n + (1 - 1/n^2)u \tag{35}$$

Here it is assumed that the medium is moving with uniform velocity. Fizeau[36] subsequently experimentally confirmed this result. Leonhardt and Piwnicki [12,13] also considered the case of non-uniform motion and showed that a moving dielectric appears to light as an effective gravitational field. They studied the metric of the space time in terms of the refractive index and the velocity of the medium as $g_{\mu\nu} = \eta_{\mu\nu} + (1/n^2 - 1)u_\mu u_\nu$, where $u_\mu u_\nu$ is normalized to unity. This metric structure at low velocity limit and the covariant metric tensor can be written as

$$g_{\mu\nu} = \begin{pmatrix} 1/n^2 & (1-1/n^2)u/c \\ (1-1/n^2)u/c & -1 \end{pmatrix} \tag{36}$$

Now as the refractive index is a function of time or z in our framework, the metric tensor will be a function of time or z. In the low velocity limit, Leonhardt showed this metric as a flat three dimensional metric. However, in our approach, this is a time dependent function. The implications of this time dependence of the metric at the cosmological level are left to a subsequent paper.

Spavieri[14] showed that the wave function for light propagation in slowly moving media is analogous to that for quantum effects of the Aharonov-Bohm type[37,38] and involves an interaction term, the electromagnetic momentum **Q** related to the flow $u$. He proposes an Aharonov-Bohm type interference experiment for measuring the phase shift of light of frequency $\omega$ from a distant star on passing near the center of a rotating cosmic object, e.g., a spiral galaxy, compared with its light passing through a flow $u$(r) at the periphery of the galaxy. He assumes the galaxy to have a hard opaque core of radius R and, at its periphery for r > R, being surrounded by the flow $u$(r) and finds that the phase shift associated with the rotating galaxy to be

$$\Delta\phi = 2\pi R u_o(\omega/c^2)(n^2 - 1), \tag{37}$$

where he assumes that the speed of the flow generally coincides with that of the core at r = R, i.e., $u$(r = R) = $u_o$ = tangential speed of the core. The present aether model indicates that such a phase shift would also depend, through $c$ and $n$, upon the z of light being emitted from the galaxy in addition to the values of R and $u_o$ obtained from the usual astrophysical observations of the galaxy. Equation (37) can be rewritten as follows

$$\Delta\phi = 2\pi R u_o(\omega/c_o^2)\, e^{-(2/\alpha)z}\, (e^{-(2/\alpha)z} - 1) \tag{38}$$

Spavieri calculates that the **Q** for a light wave dragged by the flow gives exactly the Fresnel-Fizeau momentum, **Q** = $(\omega/c^2)(n^2 - 1)$**u.** He also shows that **Q** plays the role of a magnetic vector potential. This in fact corresponds to Dirac's result[3,4] that led him to believe the aether to be necessary and motivated Allen Rothwarf[2] to pursue this aether model.

**VII - Summary**
A model of the universe as proposed by Allen Rothwarf based upon a degenerate Fermion fluid of polarizable particle-antiparticle pairs has been here extended to predict a decelerating expansion of the universe and to derive the Tully-Fisher law describing the flat rotation curves of spiral galaxies. The estimated critical acceleration parameter, $a_{Ro}$, was compared to the experimental, critical modified Newtonian Dynamics (MOND) cosmological acceleration constant, $a_o$, obtained by fitting a large number of rotation curves. The present estimated value was found to be in closer agreement with the experimental value than that obtained with the other models.

This result for $a_R(t)$ allows the derivation of the time dependent radius of the universe R(t) as a function of red shift z, R(z). In this extended model various cosmological parameters such as the velocity of light, Hubble's constant, the Tully-Fisher relation, and the index of refraction of the aether can also be expressed in terms of z. R(z) is compared with the statistical fitting for Veron-Cetty data (2006) for quasar red shifts and good agreement is found. This model also determines the time and/or z dependence of certain electromagnetic parameters, i.e., the permittivity $\varepsilon_v(t)$; the permeability $\mu_v(t)$; and index of refraction of free space $n(t)$. These are found to be potentially useful in various cosmological theories dealing with light passing through media in motion.

**Acknowledgements -** One of the authors, Sisir Roy, gratefully acknowledges School of Computational Sciences and Center for Earth Observation and Space Research, George Mason University, Fairfax, Virginia, USA for hospitality and wide support for this work. The authors acknowledge Joydip Ghosh, CEOSR, for his help with some of the computations and the figures; and Philip E. Bloomfield of Drexel University for his fine editing assistance.

**References**
 1. Grossing, G., http://arxiv.org/abs/quant-ph/0508079
 2. Rothwarf, A., Physics Essays **11**, 444(1998).
 3. Dirac, P.A.M., Nature **168**, 906 (1951).
 4. Dirac, P.A.M., Proc. Roy. Soc., **A209,** 291(1951).
 5. VERONCAT- Veron Quasars and AGNs (2006), **HEASARC Archive.**
 6. Milgrom, M., ApJ, **270**, a-365, b-371, c-384 (1983).
 7. Begeman, K.G., Broeils, A.H., and Sanders, R.H. MNRAS, **249,** 523 (1991).
 8. Sanders, R.H., and McGaugh, S.S., arxiv.astro-ph/0204521.
 9. Puthoff, H.E., Foundations of Physics, **32,** 927-943 (2002).
10. Webb, J.K. et al, Phys. Rev. Lett., **87,** 091301 (2001).
11. Damour, T. and Dyson, F., Nucl. Phys., **B480,** 37 (1996).
12. Leonhardt, U. and Piwnicki, P., Phys. Rev. A**60**, 4301 (1999).
13. Leonhardt, U. and Piwnicki, P., Phys. Rev. Lett. **84**, 822 (2000).
14. Spavieri, G., Eur. Phys. J. D **39**, 157 (2006).
15. Spergel, D. N., et el, astro-ph/0603440; ApJS, 148, 175 (2003).
16. Bonanos, A.Z., et al, astro-ph/0606279.
17. Sandage, A., et al ApJ, 653:843 (2006).
18. Freedman, W.L., et al ApJ 553:47 (2001)
19. Milgrom, M, Physics Letts.**A,253**, 273(1999).
20. Sanders, R,H., astro-ph/010658(2001).
21. Nusser *A.,* astro-ph/0109016(2001).
22. Beckenstein, J.D., Phys. Rev. D **70**, 083509 (2004).
23. Carmeli, M., *Cosmological Special Relativity: The Large-Scale Structure of Space, Time and Velocity*, World Scientific, Singapore (2002).
24. Carmeli, M., Int. J. Theor. Phys. **37**, 2621 (1998).
25. Carmeli, M., Int. J. Theor. Phys. **38**, 1993 (1999); **39**, 1397 (2000).
26. Carmeli, M., Int. J. Theor. Phys. **39**, 1397 (2000)
27. Hartnett, J. G., Int. J. Theor. Phys., **44,**349 (2005).
28. Hartnett, J. G., Int. J. Theor. Phys., **44**,485 (2005).
29. Hartnett, J. G., astro-ph/0501526 v6 (2005).
30. Carmeli, M., Hartnett, J.G., and Oliveira, F.J., Found. Phys. Lett.**19**, 277(2006).
31. Schwarzschild, B. Physics Today, **58**, 19 (2005).
32. Roy, S., et al – preprint (2007).
33. Vishwakarma, R.G., MNRAS, **345**, 545 (2003).
34. Goldstein, H., *Classical Mechanics,* 2ed (Addison-Wesley, Reading, Massachusetts, 1980).
35. Fresnel, A.J., Ann. Chim., Phys. **9**, 57 (1818).
36. Fizeau, H., C.R. Acad. Sci. (Paris) **33**, 349 (1851).
37. Aharonov, Y., Bohm, D., Phys. Rev. **115**, 485 (1959).
38. Aharonov, Y., Casher, A., Phys. Rev. Lett. **53**, 319 (1984).

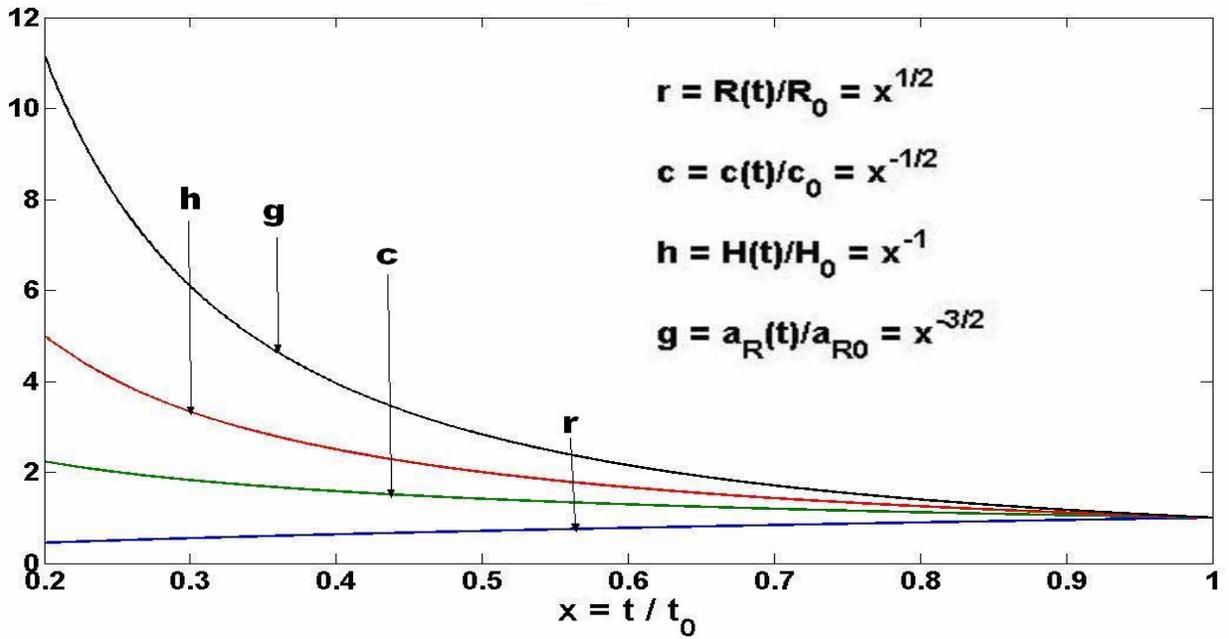

Figure 1. Reduced variables for the radius of the universe, r; the velocity of light, c; Hubble's constant, h; and the aether acceleration parameter, g as a function of reduced cosmological time, t, where $t_o$ is the present age of the universe.

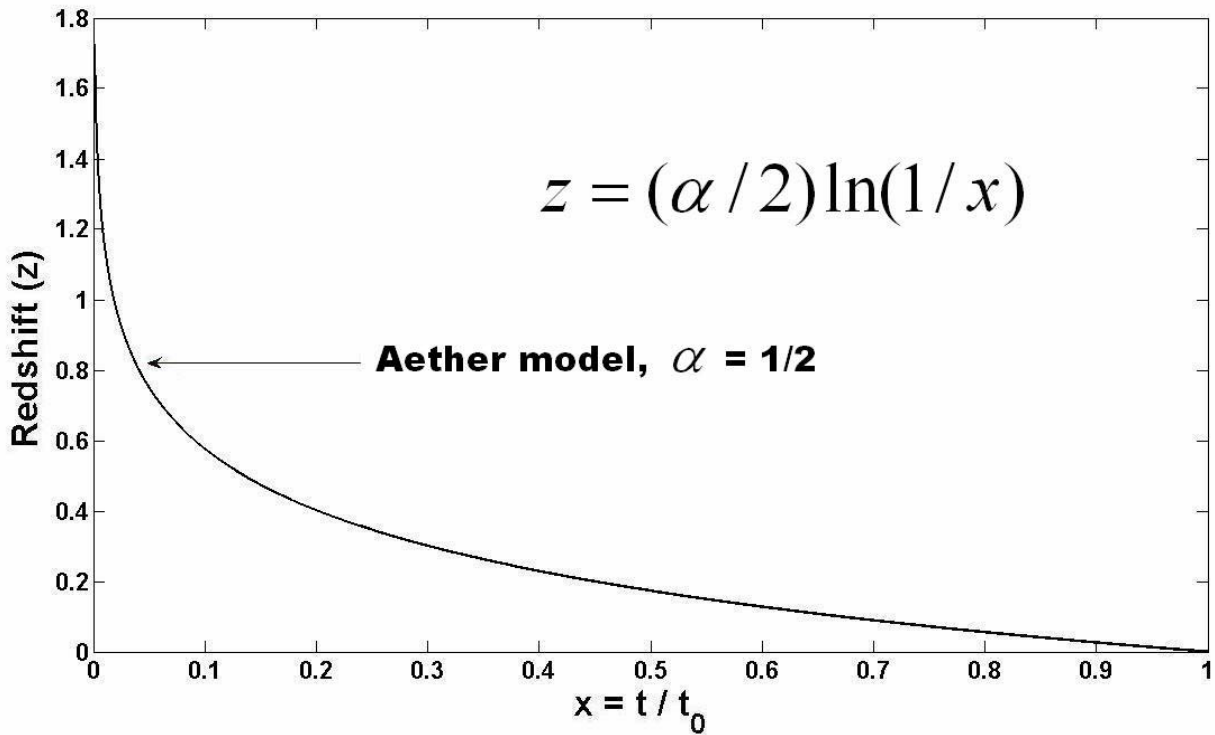

Figure 2. Redshift as a function of reduced cosmological time, **t**, for $\alpha = \frac{1}{2}$.

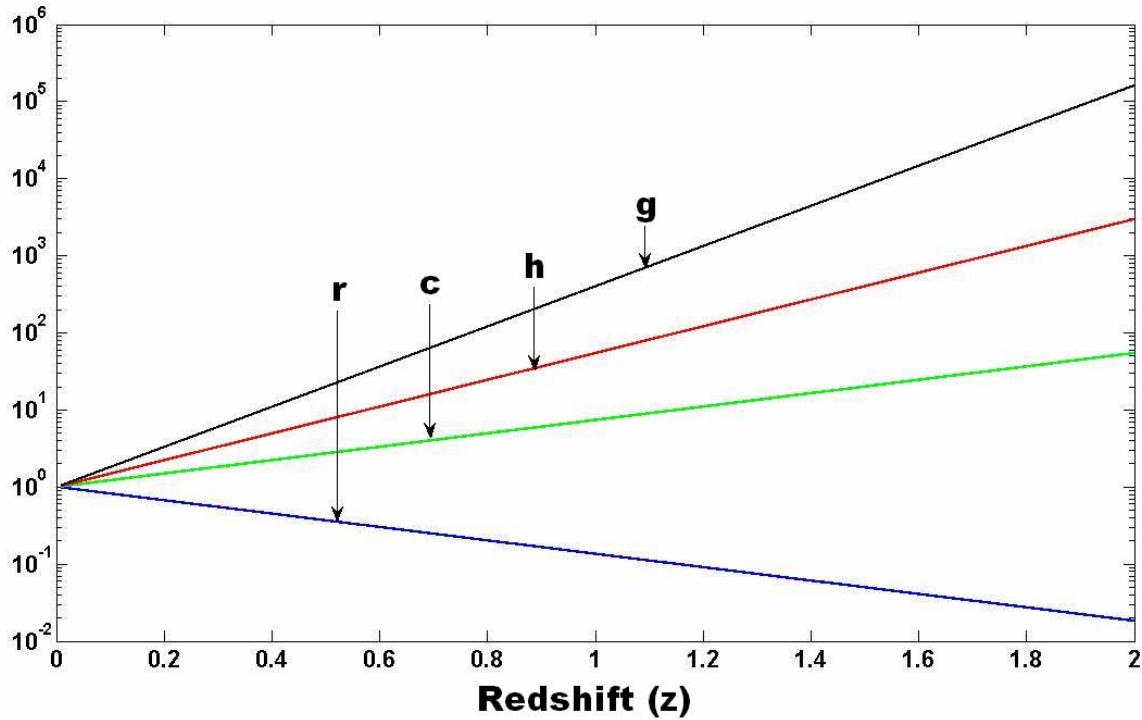

Figure 3. Logarithm of reduced variables **r, c, h, and g** as a function of redshift, z, for $\alpha = \frac{1}{2}$.

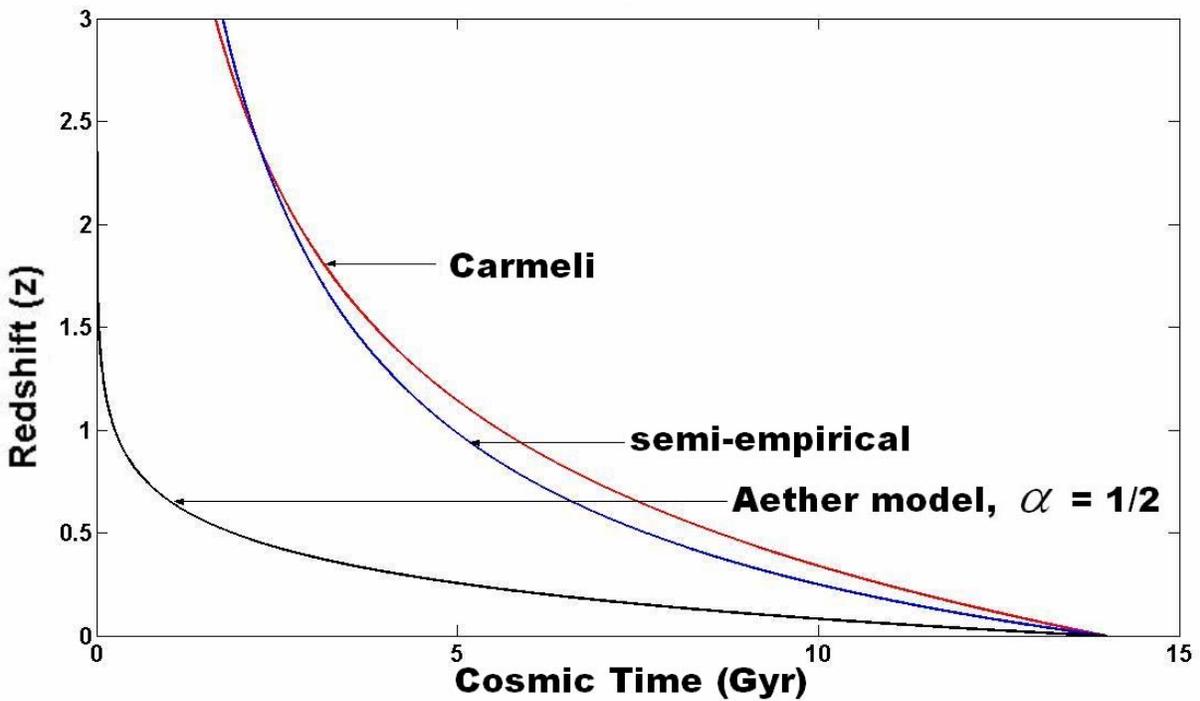

Figure 4. Redshift, z, as a function of cosmic time where $t_o = 14$ Gyr for the Carmeli, semi-empirical, and aether models.

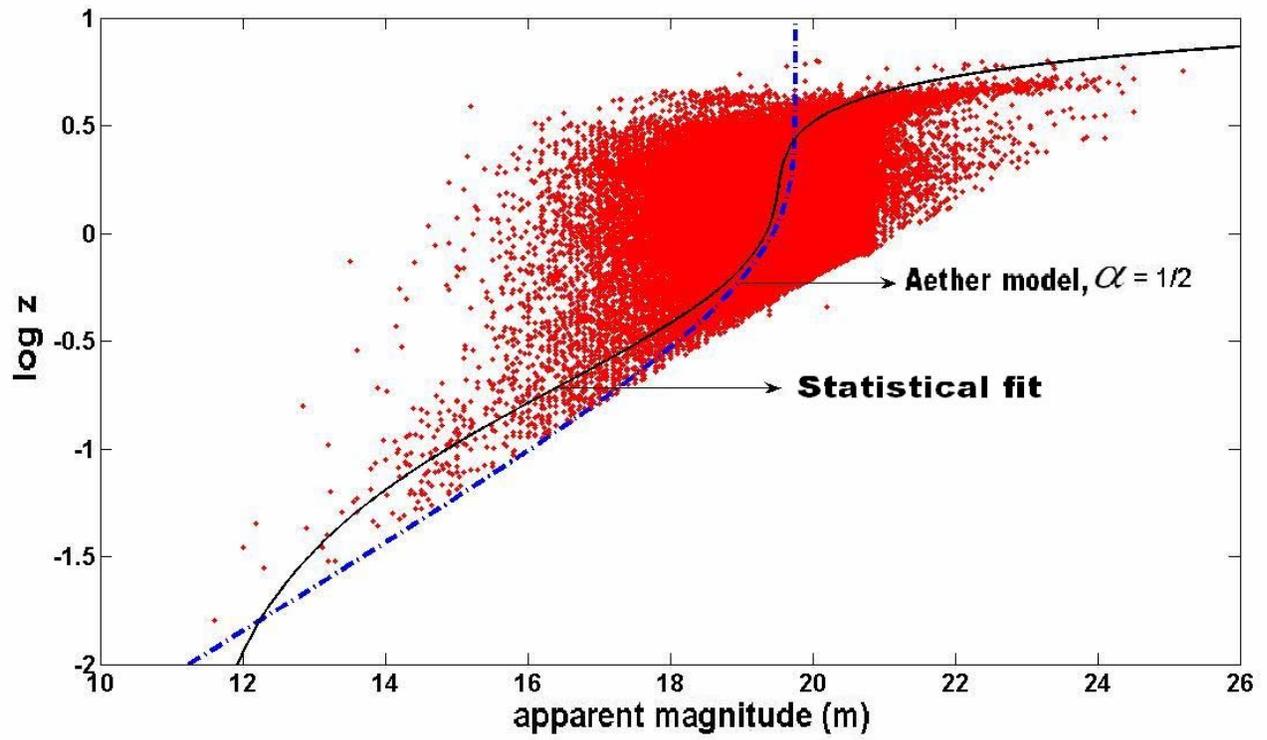

Figure 5. Statistical fit to the Veron-Cetty (2006) data for quasar redshift, z, versus apparent magnitude, **m**, compared to the prediction given by the aether model for $\alpha = \frac{1}{2}$.